\documentclass[prl,twocolumn,superscriptaddress,nofootinbib]{revtex4-2}

\usepackage{graphicx} 
\usepackage{amsmath,fullpage}
\usepackage{amssymb}
\usepackage{graphics,xcolor}
\usepackage{mathtools}
\usepackage{dcolumn}
\usepackage{bm}
\usepackage{cancel}
\usepackage[margin=2cm]{geometry}  
\usepackage{hyperref}

\def \be {\begin{equation}}
\def \ee {\end{equation}}
\def \bea {\begin{eqnarray}}
\def \eea {\end{eqnarray}}

\makeatletter 
    
\renewcommand\onecolumngrid{
\do@columngrid{one}{\@ne}%
\def\set@footnotewidth{\onecolumngrid}
\def\footnoterule{\kern-6pt\hrule width 1.5in\kern6pt}%
}

\begin{document}

\title{Violation of the third law of black hole mechanics in vacuum gravity}
\author{John R. V. Crump}
\email{jrvc2@cam.ac.uk}
\author{Maxime Gadioux}
\email{mjhg2@cam.ac.uk}
\author{Harvey~S.~Reall}
\email{hsr1000@cam.ac.uk}
\author{Jorge~E.~Santos}
\email{jss55@cam.ac.uk}
\affiliation{Department of Applied Mathematics and Theoretical Physics, University of Cambridge, Wilberforce Road, Cambridge CB3 0WA, United Kingdom}

\begin{abstract}
We demonstrate numerically the existence of solutions of five-dimensional vacuum gravity describing the formation, in finite time, of an extremal rotating black hole from a pre-existing Schwarzschild black hole. This is the first example of a violation of the third law of black hole mechanics in vacuum gravity and demonstrates that the third law is false independently of any matter model. 
We also demonstrate the existence of solutions describing the formation, in finite time, of an extremal rotating black hole from vacuum initial data that does not contain a black hole.

\end{abstract}

\maketitle

{\it Introduction.} In classical General Relativity with physically reasonable matter, the third law of black hole mechanics asserts that it is impossible for a non-extremal black hole to become extremal in finite time \cite{Bardeen:1973gs,Israel:1986gqz}. There exist counterexamples to this statement involving thin shells of charged matter \cite{shells}. However, it was argued that the distributional nature of such matter is unphysical and that the third law holds for sufficiently smooth matter fields \cite{Israel:1986gqz}. Recently, Kehle and Unger have identified a mistake in this argument and constructed counterexamples of arbitrarily high differentiability \cite{Kehle:2022uvc}. For Einstein-Maxwell theory coupled to a massless charged scalar field (or massless charged Vlasov matter \cite{Kehle:2024vyt}) their counterexamples describe spherically symmetric gravitational collapse to form an exactly extremal Reissner-Nordstr\"om black hole in finite time, with an intermediate period in which the black hole is exactly Schwarzschild at the horizon. 

These counterexamples involve matter fields with very large charge to mass ratio. It is known that similar counterexamples cannot exist if the charge to mass ratio is suitably bounded \cite{Reall:2024njy}. This raises the possibility that the third law might be true for some types of matter. However, Kehle and Unger have conjectured that there exist third-law violating solutions of {\it vacuum} gravity in which an extremal Kerr black hole forms in finite time \cite{Kehle:2022uvc,Kehle:2023eni}. If correct, this implies that the third law is false independently of any choice of matter model. But such solutions would be cohomogeneity-4 (i.e. depend on $4$ coordinates) and so constructing them is much more difficult than for the cohomogeneity-2 solutions of \cite{Kehle:2022uvc}.

We will show that third law violating solutions do exist in vacuum gravity in {\it five} spacetime dimensions. The reason for considering 5d is that there exist rotating black holes that are more symmetrical than a Kerr black hole. These are Myers-Perry black holes \cite{Myers:1986un} with two equal rotation parameters. They are cohomogeneity-1 with isometry group $\mathbb{R} \times SU(2) \times U(1)$. By combining the approach of \cite{Kehle:2022uvc} with numerical methods we will construct cohomogeneity-2 vacuum solutions which describe either (i) a Schwarzschild black hole absorbing gravitational waves to become an extremal Myers-Perry (EMP) black hole in finite time; (ii) formation of an EMP black hole in finite time from non-black hole initial data (i.e. gravitational waves). Our solutions of type (i) demonstrate that the third law is violated in 5d vacuum gravity. 

{\it The Ansatz.} There exist dynamical vacuum solutions with $SU(2) \times U(1)$ symmetry \cite{Bizon:2005cp,Dafermos:2005nh} but conservation of the Komar angular momentum defined by the $U(1)$ generator implies that they are not type (i) or (ii). We will consider spacetimes in which $U(1)$ is broken to $\mathbb{Z}_4$. Consider the group manifold of $SU(2)$, which is topologically $S^3$. $SU(2)$ acts on this group manifold via either a left action or a right action. We introduce the standard set of 1-forms $\sigma_i$ invariant under the left action. These are given in terms of Euler angles $(\theta,\phi,\psi)$ in the Supplemental Material (SM). Now consider $\mathbb{Z}_4 \subset SU(2)$ which acts via {\it right} action as $\psi \rightarrow \psi + \pi$ so $\sigma_1 \rightarrow  -\sigma_1$, $\sigma_2 \rightarrow -\sigma_2$, $\sigma_3 \rightarrow \sigma_3$. We now consider a 5d spacetime with coordinates $(U,V,\theta,\phi,\psi)$ and an $SU(2) \times \mathbb{Z}_4$ invariant metric:
 \begin{multline}
{\rm d}s^2=-\Omega^2{\rm d}U\,{\rm d}V+\frac{r^2}{4}e^{\mathcal{B}}\Big[\sqrt{1+|\Phi|^2}\,\sigma_+ \sigma_-
\\
+\frac{\overline{\Phi}}{2}\sigma_+^2+\frac{\Phi}{2} \sigma_-^2+e^{-3\mathcal{B}}\left(\sigma_3+2 \mathcal{A}\right)^2\Big]
\label{eq:doublenull}
\end{multline}
where $(U,V)$ are double null coordinates and $\sigma_\pm = \sigma_1 \pm {\rm i}\,\sigma_2$. In this Ansatz, $\Omega$, $r$, and $\mathcal{B}$ are real functions of $(U,V)$, $\Phi$ is a complex function of $(U,V)$ and ${\cal A}={\cal A}_U (U,V){\rm d}U + {\cal A}_V (U,V) {\rm d}V$ is a real 1-form. A diffeomorphism $\psi \rightarrow \psi + 2\Lambda (U,V)$ gives $\Phi \rightarrow e^{-4i\Lambda} \Phi$ and $\mathcal{A} \rightarrow \mathcal{A} + {\rm d}\Lambda$ so $\Phi$ transforms as a charged scalar field and ${\cal A}$ as a $U(1)$ gauge field. Henceforth we will use the gauge choice ${\cal A}_V=0$. The equations of motion resulting from this Ansatz are given in the SM, where we also discuss the form of the MP solution.

{\it Quasilocal mass and angular momentum.} We define the Hawking mass as \cite{Dafermos:2005nh}
\begin{equation}
m_{\rm H}(U,V)=\frac{3\pi}{8G_5}r^2\left(1+4\frac{\partial_Ur\partial_Vr}{\Omega^2}\right)\,.
\label{eq:hm}
\end{equation}
where $G_5$ is the 5d Newton constant. This satisfies $\partial_U m_{\rm H} \le 0$, $\partial_V m_{\rm H} \ge 0$ if $\partial_V r \ge 0$, $\partial_U r \le 0$ (see SM). For ${\bf m} = \partial/\partial \psi$ and $S$ a sphere of constant $U,V$ we define the angular momentum $\mathcal{J}(U,V) = \frac{1}{16\pi G_5}\int_S \star {\rm d}{\bf m}^\flat=\frac{\pi}{4 G_5} J$ where $J = (1/2) r^5 e^{-2\mathcal{B}}\Omega^{-2} \partial_V \mathcal{A}_U$. The EMP solution has $m_{\rm H}=\frac{3\pi}{8 G_5}r_+^2$ for a horizon cross-section and $J = J_{\rm EMP} \equiv r_+^3$ everywhere. 

\begin{figure*}[t]
    \centering
    \includegraphics[width=\linewidth]{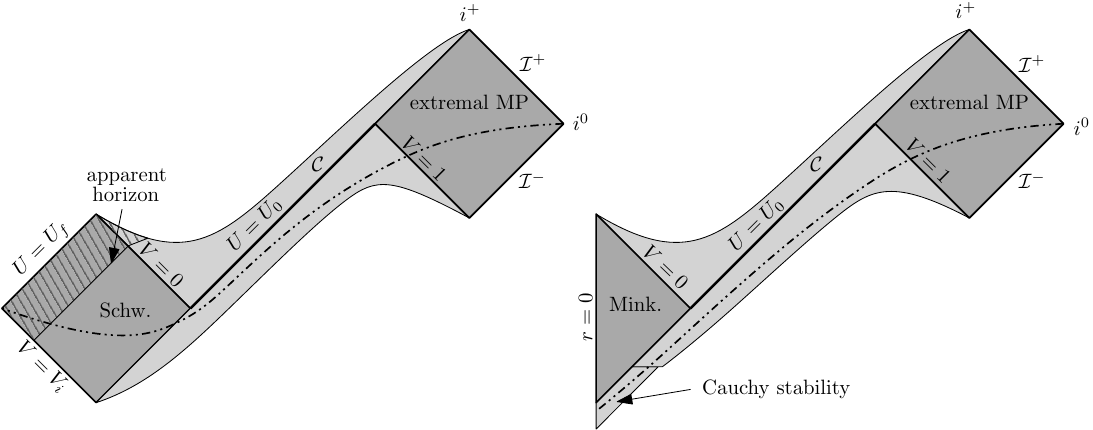}
    \caption{Penrose diagrams for gluing constructions of type (i) (left) and (ii) (right). The dark regions are exactly isometric to subsets of the Schwarzschild, EMP and Minkowski spacetimes. Light-shaded regions are local solutions obtained by solving characteristic initial/final value problems. The dash-dotted curves are sample Cauchy surfaces. In the left panel, the Schwarzschild region occupies $[U_0,U_f]\times [V_i,0]$; we do not extend it to the singularity at $r=0$. The striped pattern denotes the trapped region, its boundary is the apparent horizon. In the right panel, Cauchy stability ensures the solution in the light shaded region can be extended to the centre of symmetry $r=0$ using an argument in \cite{Kehle:2022uvc}.}
    \label{fig:penrose}
\end{figure*}

{\it Characteristic gluing.} We use the method of characteristic gluing \cite{gluing} as in \cite{Kehle:2022uvc,Gadioux:2025unn}. We will prescribe data on a null hypersurface $\mathcal{C}$ with equation $U=U_0$ that for $V\le 0$ coincides exactly with the data of an outgoing spherically symmetric null hypersurface outside the apparent horizon in a Schwarzschild spacetime and for $V\ge 1$ coincides exactly with the data on the event horizon of EMP. See Fig.~\ref{fig:penrose}, left panel. Given such data, we consider a spacetime that is exactly Schwarzschild in a region $U>U_0$, $V<0$. The characteristic initial value problem defined by the data on the surfaces $\{U=U_0,V\ge 0\}$ and $\{V = 0,U \ge U_0\}$ admits a local solution \cite{characteristic}, so our spacetime can be extended, as a solution of the Einstein equation, into a region to the future of $\mathcal{C}$, shown in light grey. We next assume that the spacetime is exactly EMP in the region $U \le U_0$, $V \ge 1$. The spacetime can be extended into the light grey region to the past of $\mathcal{C}$ by solving the characteristic final value problem defined by the data on the surfaces $\{U=U_0,V \le 1\}$ and $\{V=1,U \le U_0\}$. The problem is to choose the data for $V \in (0,1)$ on $\mathcal{C}$ such that the resulting spacetime is suitably smooth across $\mathcal{C}$. If it is then we have constructed a solution of type (i). For type (ii) we replace the early time Schwarzschild data with data on an outgoing spherically symmetric null cone in Minkowski spacetime as shown in Fig.~\ref{fig:penrose}, right panel.

We aim to construct a solution with a metric that is $C^k$ on $\mathcal{C}$. The equations of motion then imply that $J$ is $C^k$ and $r$ is $C^{k+1}$. We have the gauge freedom
\begin{equation}
\label{eq:gauge}
 U' = f(U) \qquad V' = g(V) \qquad \Omega^{'2} = \frac{\Omega^2}{f'(U) g'(V)}\,.
\end{equation}
We use this to set $\Omega(U_0,V) \equiv 1$ and $\partial_U^j \Omega(U_0,0)=0$ ($1 \le j \le k$). The free data are the apparent horizon radius $r_i$ of the initial Schwarzschild solution, the horizon radius $r_+$ of the final EMP solution and the fields $\mathcal{B}$ and $\Phi$ on $\mathcal{C}$. From these, $r$ is determined on $\mathcal{C}$ by solving \eqref{eq:rVV} backwards starting with $r=r_+$ and $\partial_V r=0$ at $V=1$. (Gluing at $V=1$ removes the residual freedom $g'(V) = 1/f'(U_0)$ in \eqref{eq:gauge}.) The data must be chosen such that $r(0)>0$. $r_i$ is then chosen such that $r(0)>r_i>0$. Equation \eqref{eq:JV} is used to propagate $J$ along $\mathcal{C}$. The data must be chosen so that $J$ matches $J_{\rm EMP}$ at $V=1$. The initial Schwarzschild hole has $m_{\rm H} \equiv 3\pi r_i^2/(8 G_5)$ so \eqref{eq:hm} determines $\partial_U r(0)$, then \eqref{eq:rUV} can be used to propagate $\partial_U r$ along $\mathcal{C}$. We require that $\partial_U r<0$ along $\mathcal{C}$ to ensure that the gluing does not occur inside a white hole. To check smoothness on $\mathcal{C}$ note that transverse derivatives $\partial^j_U \Omega$, $\partial^j_U r$, $\partial_U^j \Phi$, $\partial_U^j \mathcal{B}$, $\partial_U^j \mathcal{A}_U$ on $\mathcal{C}$ satisfy first order transport equations on $\mathcal{C}$ obtained by taking $U$-derivatives of equations of motion. Hence these quantities are  uniquely defined by their values at either $V=0$ or $V=1$. We need these two choices to give the same result, i.e., when transported along $\mathcal{C}$ starting at $V=0$ these quantities must match their values at $V=1$. We can apply \eqref{eq:gauge} with $g'(V)=1/f'(U_0)$ to the EMP solution (in $\Omega(U_0,V) \equiv 1$ gauge), and by choosing $f'(U_0)$ and $f^{(j+1)}(U_0)$ appropriately, this can be used to ensure that the EMP values of $\partial_U r$ and $\partial_U^j \Omega$($1\le j \le k$) match our solution at $V=1$. Similarly, the residual gauge symmetry $\psi \rightarrow \psi + 2\Lambda(U)$ can be used to ensure that the EMP values of $\partial^j_U \mathcal{A}_U$ ($0 \le j \le k$) match our solution at $V=1$. Constraints on the free data arise from matching of $\partial_U^j \Phi$ and $\partial_U^j \mathcal{B}$ ($j \le k$). At $V=1$ we need $\partial_U^j \Phi$ to match its EMP value (zero). To match $\partial_U^j \mathcal{B}$ we define $L \equiv \Omega^{-2} \partial_U$. The quantities $\mathcal{B}^j = [(L^j \mathcal{B})/(L r)^j]$ are gauge invariant under \eqref{eq:gauge} and so at $V=1$ need to match their EMP values given in the SM. Equation \eqref{eq:rUU} then implies that $\partial_U^j r$ ($2 \le j \le k+1$) must also match its EMP value. (We use this as a check on our numerical solutions below, finding a relative error less than $10^{-21}$.)
In summary, the problem is to choose the free data to achieve the matching of $J$, $\mathcal{B}^j$ and $\partial_U^j \Phi$ ($0 \le j \le k$) whilst ensuring $r>0$ and $\partial_U r<0$ for $V \in [0,1]$.

Solutions of type (i) constructed this way are $C^k$ everywhere (not just on $\mathcal{C}$). Solutions of type (ii) extend to $r=0$, where the equations are singular. As in \cite{Kehle:2022uvc} there is a loss of regularity at $r=0$. In the SM we show that, away from $r=0$, our solutions will be piecewise $C^{k+1}$ with bounded $(k+1)$th derivatives and that this implies that they will be $C^{k-2}$ at $r=0$. So achieving a type (ii) solution that is everywhere $C^2$ requires $k=4$.

{\it Numerical approach.} Our Ansatz for the gluing functions on $[0,1]$ is
\begin{align}
    \Phi(U_0,V) &= \varphi(V) P(V)  \, e^{i \omega(V) V} \label{eq:PhiAnsatz} \\
    \mathcal{B}(U_0,V) &= \beta(V)P(V) - \frac{1}{3}\log 2 \, Q(V) \label{eq:BAnsatz}
\end{align}
\noindent where $P,Q$ are explicit real functions (see SM) that are $C^4$, piecewise $C^5$, with $P(0)=P(1)=Q(0)=0$, $Q(1)=1$ and the first $4$ derivatives of $P,Q$ vanish at $V=0,1$. $\varphi$, $\omega$ and $\beta$ are real functions each taking the form of a single hidden layer neural network with activation function $\tanh$:
\begin{equation}
    \chi_0 + \sum_{i=1}^{p} \chi_i \tanh{(\mu_i V+\kappa_i)}\,.
\end{equation}
The values of $p$, and the parameters $\chi_0$, $\chi_i$, $\mu_i$, $\kappa_i$ may differ for $\varphi$, $\omega$ and $\beta$. We denote the set of all parameters as $\boldsymbol{\theta}$.
We perform the gluing procedure numerically. A weighted sum of the squared residuals of the gluing conditions is used as an objective function $L(\boldsymbol{\theta})$. A set of parameters $\boldsymbol{\theta}$ provides a solution when $L(\boldsymbol{\theta})=0$, $r>0$ and $\partial_Ur<0$. At initialization, $\chi_i$ and $\mu_i$ are randomly sampled from $\mathcal{N}(0,1)$, $\kappa_i=-\frac{1}{2}\mu_i$, and $\chi_0$ is set to zero. We repeatedly sample $\chi_i$ and $\mu_i$, choosing the set with the least $L(\boldsymbol{\theta})$ as the seed. Gradient descent with the Adam optimizer \cite{Kingma:2015} is used to update the parameters, where the gradient is approximated by a fourth-order central difference formula:
\begin{equation}
\begin{split}
    (\nabla_{\boldsymbol{\theta}} L)_i \approx \frac{1}{12 \epsilon} \Big[ 
& L(\boldsymbol{\theta} - 2\epsilon \boldsymbol{e}_i) 
- 8 L(\boldsymbol{\theta} - \epsilon \boldsymbol{e}_i) \\
& + 8 L(\boldsymbol{\theta} + \epsilon \boldsymbol{e}_i) 
- L(\boldsymbol{\theta} + 2\epsilon \boldsymbol{e}_i) 
\Big] + \mathcal{O}(\epsilon^4)\,. 
\end{split}
\end{equation}
In this stage, penalty terms are added to $L(\boldsymbol{\theta})$ to prevent $r$ and $\partial_Ur$ from vanishing. When $L(\boldsymbol{\theta})$ becomes small, the penalty terms are removed and the BFGS algorithm \cite{BFGS:1970} is used to reach a local minimum. If this does not exhibit convergence to zero, the $p$ values are increased. Typically, convergence to a solution from a random seed was faster with a large number of parameters. Solutions with fewer parameters were then found by using a fewer-parameter approximation to the higher-parameter solution as a seed. Once a good candidate solution has been identified, we switch to a traditional quasi-Newton method implemented in extended precision. In this final stage, the quasi-Newton search is applied only to a subset of the parameters, chosen so that its dimension matches the number of gluing conditions to be imposed. Specifically, we search for solutions for which $\mathcal{B}^j$, $\partial_U^j \Phi$ (with $1 \le j \le k$), and $J$ are fixed to their EMP values, while $r(0)$ and $\partial_U r(1)$ are fixed to the values obtained in the final iteration of the BFGS algorithm. This yields $3k+3$ conditions, which are satisfied by tuning $3k+3$ parameters in the final Newton solve, initialised from the last BFGS iterate. We present extensive convergence tests in the SM.

\begin{figure*}[t]
    \centering
    \includegraphics[width=\linewidth]{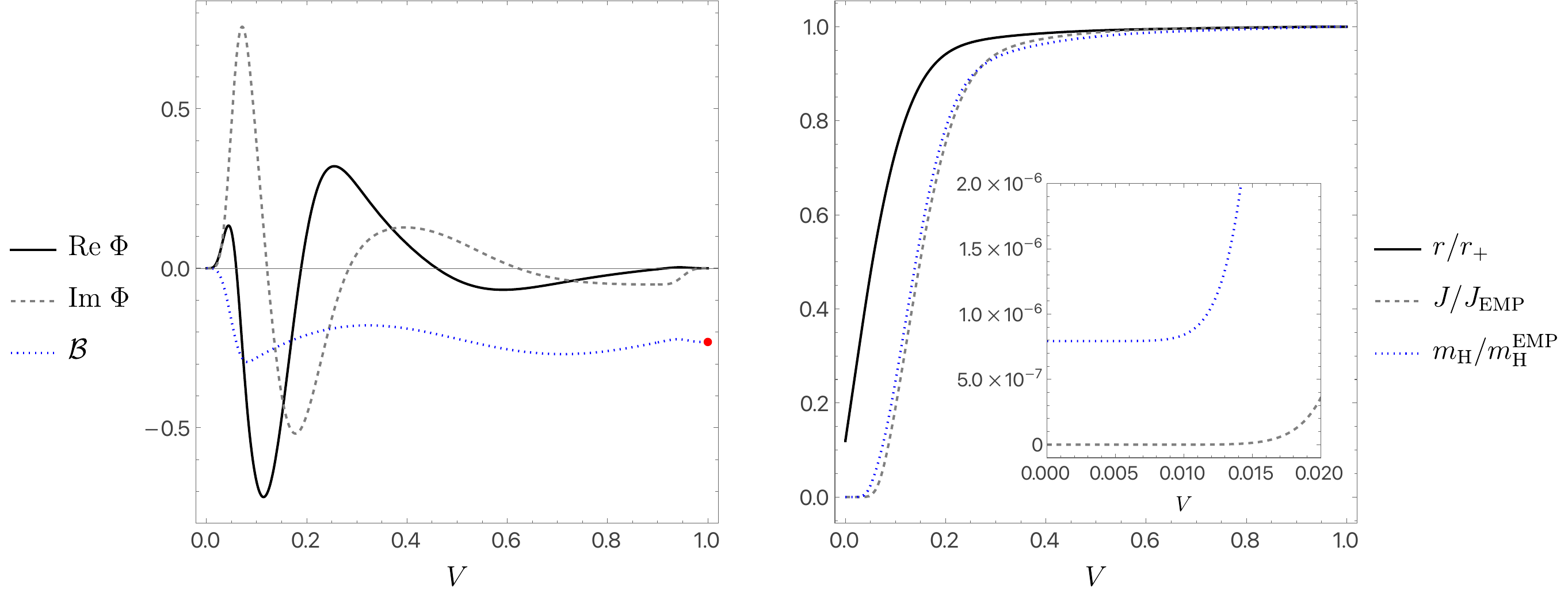}
    \caption{A $k=2$ gluing solution of type (i). Left: $\Phi$ and $\mathcal{B}$, with the red dot marking the EMP limit of $\mathcal{B}$. Right: $r$, $J$ and $m_{\rm H}$. The inset plot on the right panel shows that $J=0$ but $m_{\rm H}>0$ at $V=0$.}

\label{fig:type_I}
\end{figure*}
\begin{figure*}[t]
    \includegraphics[width=\linewidth]{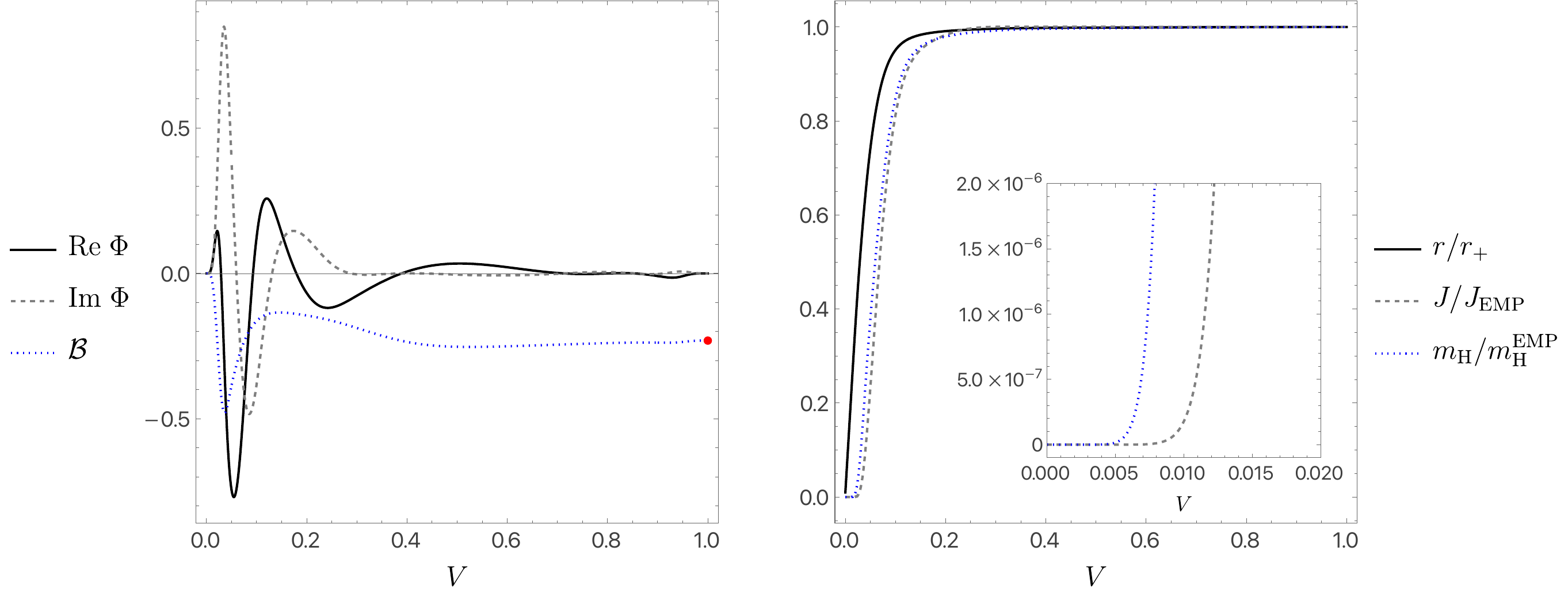}
    \caption{A $k=4$ gluing solution of type (ii).} \label{fig:type_II}
\end{figure*}

{\it Results.} We have obtained $k=2$ solutions choosing $p=(2,2,3)$ for $\varphi$, $\omega$ and $\beta$, a total of $24$ parameters in our Ansatz. With so many parameters we do not expect solutions to be unique \cite{Gadioux:2025unn}. Fig.~\ref{fig:type_I} shows the behaviour of $\Phi$, $\mathcal{B}$, $r$, $J$ and $m_{\rm H}$ on $\mathcal{C}$ for a typical solution of type (i) with $k=2$. This solution has $r_i/r_+ \approx 0.0008909$ and $r(U_0,0)/r_+ \approx 0.1194095$ so the initial event horizon lies well outside the Schwarzschild radius. $\Phi$ and $\mathcal{B}$ are largest at early time on $\mathcal{C}$. This also occurs for gluing to extremal Reissner-Nordstr\"om, which can be understood as a consequence of Raychaudhuri's equation \cite{Gadioux:2025unn}.

Type (ii) solutions with $k=2$ are no harder to find numerically but we need $k=4$ for a type (ii) solution that is $C^2$ at $r=0$. Finding $k=4$ solutions numerically is considerably more challenging because there are many more equations to solve, $\mathcal{B}^j$ increases rapidly with $j$, and $V$-derivatives of quantities are larger. To find $k=4$ solutions we needed more parameters: $p=(4,16,8)$ for $\varphi$, $\omega$ and $\beta$, a total of $87$ parameters. Fig.~\ref{fig:type_II} shows such a solution. It has $r(U_0,0)/r_+ \approx 0.0097801$, much smaller than for $k=2$. 

Plots of other quantities for our solutions are provided in the SM. Parameter values for the solutions are  available in the GitHub repository (\href{https://github.com/jorgealberich/Violation-of-the-third-law-of-black-hole-mechanics-in-vacuum-gravity}{GitHub}).

{\it Discussion.} Our solutions of type (i) describe a tiny Schwarzschild black hole absorbing gravitational waves to become a large EMP black hole in finite time, a violation of the third law of black hole mechanics in vacuum. Our solutions of type (ii) describe formation of an EMP black hole from gravitational collapse of gravitational waves. It should be possible to perform a 2-stage gluing construction, as in \cite{Kehle:2022uvc}, to combine types (i) and (ii) to construct a solution which starts from non-black hole initial data and forms an EMP black hole with an intermediate Schwarzschild phase. The method can also be used to obtain solutions describing formation of a subextremal MP black hole with any parameter values. 

Numerical methods could be used to study the properties of these solutions away from $\mathcal{C}$. For example, do type (ii) solutions contain trapped surfaces? If so, they cannot be critical solutions \cite{Kehle:2024vyt} (solutions at the threshold between black hole formation and dispersion). One could also try to construct critical solutions numerically by considering Cauchy evolution in the class of $SU(2) \times \mathbb{Z}_4$ symmetric spacetimes \eqref{eq:doublenull}.

A more ambitious goal is to perform the gluing construction for 4d vacuum gravity to construct solutions describing finite time formation of an extremal Kerr black hole.

{\it Acknowledgments.}
We thank M. Dafermos, C. Kehle and R. Unger for helpful discussions. 
JRVC and MG are supported by STFC studentships 
and the Cambridge Trust. 
HSR and JES are supported by STFC grant ST/X000664/1. JES is also supported by Hughes Hall College. Part of this work was completed when MG and HSR were in residence at ICERM, Brown University, supported by NSF Grant No. DMS-1929284, and also when MG was in residence at IFPU, Trieste.

\onecolumngrid
\section{Supplemental Material}

\subsection{Left invariant $1$-forms}

The left-invariant $1$-forms on $SU(2)$ are
\begin{equation}
\begin{aligned}
&\sigma_1 = \cos \psi\,\sin \theta\,{\rm d}\phi-\sin \psi\,{\rm d}\theta\,,
\\
&\sigma_2 = \sin \psi\,\sin \theta\, {\rm d}\phi+\cos \psi\,{\rm d}\theta\,,
\\
&\sigma_3={\rm d}\psi+\cos \theta \,{\rm d}\phi\,,
\end{aligned}
\end{equation}%
with $\theta\in[0,\pi]$, $\phi\sim \phi+2\pi$ and $\psi\sim \psi+4\pi$.

\subsection{Equations of motion}

The equations of motion resulting from our metric Ansatz are:

\begin{subequations}
\begin{equation}
\label{eq:rVV}
\frac{\Omega^2}{r}\partial_V\left(\frac{\partial_Vr}{\Omega^2}\right)=-\frac{1}{2}\left(\partial_V\mathcal{B}\right)^2-\frac{1}{6}\frac{\left|\partial_V\Phi\right|^2}{1+|\Phi|^2}-\frac{1}{24}\frac{\left|\overline{\Phi}\partial_V\Phi-\Phi \overline{\partial_V \Phi}\right|^2}{1+|\Phi|^2}\,,
\end{equation}

\begin{equation}
\label{eq:rUU}
\frac{\Omega^2}{r}\partial_U\left(\frac{\partial_Ur}{\Omega^2}\right)=-\frac{1}{2}\left(\partial_U \mathcal{B}\right)^2-\frac{1}{6}\frac{\left|\mathcal{D}_U\Phi\right|^2}{1+|\Phi|^2}-\frac{1}{24}\frac{\left|\bar{\Phi}\mathcal{D}_U\Phi-\Phi \overline{\mathcal{D}_U \Phi}\right|^2}{1+|\Phi|^2}\,,
\end{equation}

\begin{equation}
\label{eq:JV}
\partial_VJ=r^3 {\rm Im}\left(\overline{\Phi} \partial_V \Phi\right)\,,
\end{equation}
\begin{equation}
\partial_UJ=r^3 {\rm Im}\left(\Phi \overline{\mathcal{D}_U \Phi}\right)\,,
\end{equation}
\begin{equation}
\label{eq:AUV}
\partial_V\mathcal{A}_U=\frac{2J\,e^{2\mathcal{B}}\Omega^2}{r^5}\,,
\end{equation}

\begin{equation}
\label{eq:rUV}
\partial_U \partial_Vr-\frac{\Omega^2e^{-4\mathcal{B}}}{6 r}+\frac{2\,\partial_U r\,\partial_V r}{r}-\frac{2}{3}\frac{\Omega^2e^{2\mathcal{B}}J^2}{r^7}+\frac{2}{3}\frac{\Omega^2e^{-\mathcal{B}}}{r}\left(\sqrt{1+|\Phi|^2}-e^{3\mathcal{B}}|\Phi|^2\right)=0\,,
\end{equation}

\begin{multline}
\partial_U \partial_V \log\left(\Omega^2\right)+\frac{\Omega^2e^{-4\mathcal{B}}}{2 r^2}+\frac{6 J^2e^{2\mathcal{B}} \Omega^2}{r^8}-\frac{6 \partial_U r\partial_V r}{r^2}+\frac{3}{2}\partial_U \mathcal{B}\,\partial_V \mathcal{B}+\frac{1}{2}{\rm Re}\left(\mathcal{D}_U \Phi \overline{\partial_V \Phi}\right)
\\
-\frac{1}{8}\frac{\partial_U(|\Phi|^2)\partial_V(|\Phi|^2)}{1+|\Phi|^2}-\frac{2\Omega^2e^{-\mathcal{B}}}{r^2}\left(\sqrt{1+|\Phi|^2}-e^{3\mathcal{B}}|\Phi|^2\right)=0\,,
\end{multline}

\begin{equation}
\label{eq:BUV}
\partial_U \partial_V \mathcal{B}+\frac{3}{2}\frac{\partial_U \mathcal{B}\,\partial_V r}{r}+\frac{3}{2}\frac{\partial_V \mathcal{B}\,\partial_U r}{r}+\frac{4J^2e^{2\mathcal{B}}\Omega^2}{3r^8}+\frac{2}{3}\frac{\Omega^2e^{-\mathcal{B}}}{r^2}\left(\sqrt{1+|\Phi|^2}+2e^{3\mathcal{B}}|\Phi|^2-e^{-3\mathcal{B}}\right)=0\,,
\end{equation}

\begin{multline}
\label{eq:PhiUV}
\partial_U \partial_V \Phi+\frac{3}{2}\frac{\mathcal{D}_U \Phi\partial_V r}{r}+\frac{3}{2}\frac{\partial_V \Phi\partial_U r}{r}+\frac{4 {\rm i} e^{2\mathcal{B}}\Omega^2J}{r^5}\Phi+4{\rm i}\mathcal{A}_U\partial_V \Phi
\\
+\frac{1}{2}\frac{{\rm Re}\left(\overline{\Phi}^2\mathcal{D}_U \Phi\partial_V \Phi\right)}{1+|\Phi|^2}\Phi-\frac{1}{2}\frac{2+|\Phi|^2}{1+|\Phi|^2}{\rm Re}(\partial_V\Phi \overline{\mathcal{D}_U\Phi})\Phi-\frac{2\sqrt{1+|\Phi|^2}\Omega^2e^{-\mathcal{B}}}{ r^2}\Phi+\frac{4e^{2\mathcal{B}}\Omega^2}{r^2}\left(1+|\Phi|^2\right)\Phi=0\,,
\end{multline}
\end{subequations}%
where $\mathcal{D}_U\Phi\equiv\partial_U\Phi+4{\rm i}\mathcal{A}_U\Phi$.

\subsection{Monotonicity of the Hawking mass}

The equations of motion imply
\begin{subequations}
\begin{equation}
\frac{8 G_5}{3\pi}\partial_V m_{\rm H} = 
\Biggl[\frac{8 J^2 e^{2\mathcal{B}}}{3 r^5} 
    + \frac{2 r e^{-4 \mathcal{B}}}{3} f(e^{\mathcal{B}},|\Phi|^2)\Biggr] \partial_V r
+ \frac{2 r^3}{\Omega^2} \Biggl\{(\partial_V \mathcal{B})^2 + \frac{|\partial_V \Phi|^2}{3(1+|\Phi|^2)} + \frac{\left[{\rm Im}(\bar{\Phi} \partial_V \Phi)\right]^2}{3(1+|\Phi|^2)}\Biggr\} (-\partial_U r) \,,
\end{equation}
and
\begin{equation}
\frac{8 G_5}{3\pi}\partial_U m_{\rm H} = 
-\Biggl[\frac{8 J^2 e^{2\mathcal{B}}}{3 r^5} 
    + \frac{2 r e^{-4 \mathcal{B}}}{3} f(e^{\mathcal{B}},|\Phi|^2)\Biggr] (-\partial_U r)
-\frac{2 r^3}{\Omega^2} \Biggl\{(\partial_U \mathcal{B})^2 + \frac{|\mathcal{D}_U \Phi|^2}{3(1+|\Phi|^2)} + \frac{\left[{\rm Im}(\bar{\Phi} \mathcal{D}_U \Phi)\right]^2}{3(1+|\Phi|^2)}\Biggr\} \partial_V r \,,
\end{equation}
with 
\begin{equation}
f(x,y)=1+3x^4+4x^6y-4x^3\sqrt{1+y}\,.
\end{equation}
\end{subequations}%
(We note that $f= 3e^{4\mathcal{B}} (1-r^2 \mathcal{R}/6)$ where $\mathcal{R}$ is the Ricci scalar of a homogeneously deformed $S^3$ of constant $U,V$.) We claim that $f(x,y)\geq0$ for $x>0$ and $y\geq0$. To prove this we fix $x$ and consider $\partial_y f = 2x^3 (2x^3 - 1/\sqrt{1+y})$. For $x^3 \ge 1/2$ this gives $\partial_y f \ge 0$ so, along a line of constant $x$, $f$ is minimized at $y=0$ where we have $f(x,0)=(1-x)^2 \left(1+2 x+3 x^2\right) \ge 0$. For $0<x^3<1/2$, $\partial_y f<0$ for $y<y_{\rm min}$ and $\partial_y f>0$ for $y>y_{\rm min}$ where $y_{\rm min} = 1/(4x^6)-1$. Hence, along a line of constant $x$, $f$ is minimized at $y=y_{\rm min}$ where we have $f(x,y_{\rm min})=x^4 \left(3-4 x^2\right)$ which is positive for $0<x^3<1/2$. This completes the proof. Note that $f(1,0)=0$ so $x=1$, $y=0$ is the global minimum of $f$, which corresponds to a round sphere. 

It now follows that $\partial_U m_{\rm H} \le 0$ and $\partial_V m_{\rm H} \ge 0$ if $\partial_V r \ge 0$ and $\partial_U r \le 0$. For $\Phi=0$ this is a special case of \cite{Dafermos:2005nh}.



\subsection{The Myers-Perry Black Hole}

The equal-angular-momenta Myers–Perry black hole, in Boyer–Lindquist coordinates $(t,\tilde{r},\theta,\phi,\tilde{\psi})$, is given by
\begin{subequations}
\label{MPoriginal}
\begin{equation}
{\rm d}s^2=-\frac{f(\tilde{r})}{h(\tilde{r})}{\rm d}t^2+\frac{{\rm d}\tilde{r}^2}{f(\tilde{r})}+\frac{\tilde{r}^2}{4}\left\{{\rm d}\theta^2+\sin^2\theta{\rm d}\phi^2+h(\tilde{r})\left[{\rm d}\tilde{\psi}+\cos\theta\,{\rm d}\phi-2 w(\tilde{r}){\rm d}t\right]^2\right\}
\end{equation}
where $(\theta,\phi,\tilde{\psi})$ are Euler angles on $S^3$ and
\begin{equation}
h(\tilde{r})=1+\frac{\tilde{r}_0^2 a^2}{\tilde{r}^4}\,,\quad f(\tilde{r})=1+\frac{\tilde{r}_0^2 a^2}{\tilde{r}^4}-\frac{\tilde{r}_0^2}{\tilde{r}^2}\quad\text{and}\quad w(\tilde{r})=\frac{a \tilde{r}_0^2}{\tilde{r}^4h(\tilde{r})}\,.
\end{equation}
\end{subequations}%
Note that $\tilde{r}_0$ enters the metric quadratically, so we may take $\tilde{r}_0>0$ without loss of generality. As long as $|a|\leq\tilde{r}_0/2$, $f(\tilde{r})$ has two positive real roots. The location of the event horizon is given by the largest root $\tilde{r}_+$ which is determined by
\begin{equation}\label{r0FUNCTIONrpt}
\tilde{r}_0=\frac{\tilde{r}_+^2}{\sqrt{\tilde{r}_+^2-a^2}}\,.
\end{equation}
The extremal MP solution has $|a|=\tilde{r}_0/2=\tilde{r}_+/\sqrt{2}$. It has a smooth event horizon. One can convert to coordinates regular at the horizon using standard methods. This involves a shift $\psi = \tilde{\psi}+\ldots$ where the ellipsis is a certain function of $\tilde{r}$ that diverges at $\tilde{r}=\tilde{r}_+$. We then have ${\bf m} = \partial/\partial \psi = \partial/\partial \tilde{\psi}$, which is a Killing vector field so our quasi-local angular momentum is constant in spacetime and agrees with the Komar angular momentum. This gives
\be
 J = \frac{\alpha\,\tilde{r}_+^3}{1-\alpha^2}
\ee
where $\alpha\equiv a/\tilde{r}_+$ (so $|\alpha| = 1/\sqrt{2}$ at extremality).

We can compare the geometry of the $S^3$ orbits of the $SU(2)$ symmetry with \eqref{eq:doublenull} to read off
\begin{equation}
    r=\tilde{r} h(\tilde{r})^{1/6}\qquad  \qquad \Phi=0 \qquad \qquad e^{\mathcal{B}} = \frac{\tilde{r}^2}{r^2}
\end{equation}
The first equation implies
\begin{equation}
r_+=\frac{{\tilde{r}_+}}{(1-\alpha ^2)^{1/6}}\,.
\end{equation}
To define double null coordinates we can introduce a tortoise coordinate $\tilde{r}_\star$ such that ${\rm d}\tilde{r}_{\star}=h^{1/2}f^{-1}{\rm d}r$ and then define retarded and advanced time coordinates $u=t-r_\star$, $v=t+r_\star$. The metric now has exactly the form \eqref{eq:doublenull} with $U=u$, $V=v$ and $\Omega^2 = f/h$. A change of double null gauge \eqref{eq:gauge} is still required to obtain coordinates regular at the horizon. However, all we need to know is the quantities $\mathcal{B}^j$ defined in the main text, and these are defined to be gauge invariant under \eqref{eq:gauge}. Hence we can calculate them in $(u,v)$ coordinates outside the horizon and then take a limit $u \rightarrow \infty$ at fixed $v$ to obtain their values on the future event horizon. This is straightforward because $\partial_u$ can be expressed in terms of $\partial_t$ and $\partial_{\tilde{r}}$. The result is
\begin{multline}
\mathcal{B}^0=\frac{1}{3} \log \left(1-\alpha ^2\right)\,,\quad \tilde{r}_+\mathcal{B}^1=\frac{4 \alpha ^2 (1-\alpha ^2)^{1/6}}{3-2 \alpha ^2}\,,\quad \tilde{r}^2_+\mathcal{B}^2=-\frac{12 \alpha ^2 \left(5-2 \alpha ^2\right) (1-\alpha ^2)^{1/3}}{\left(3-2 \alpha ^2\right)^2}\,,
\\
\tilde{r}_+^3\mathcal{B}^3=\frac{72 \alpha ^2 (1-\alpha ^2)^{1/2} \left(15-10 \alpha ^2+4 \alpha ^4\right)}{\left(3-2 \alpha ^2\right)^3}\quad\text{and}\quad \tilde{r}_+^4\mathcal{B}^4=-\frac{216 \alpha ^2 \left(1-\alpha ^2\right)^{2/3} \left(105-80 \alpha ^2+80 \alpha ^4-24 \alpha ^6\right)}{\left(3-2 \alpha ^2\right)^4}\,.
\end{multline}

\subsection{Regularity at the origin}

For solutions of type (ii), consider data on $\mathcal{C}$ that is $C^k$ and piecewise $C^{k+1}$ with bounded discontinuities in $(k+1)$th $V$-derivatives at $V=0,\delta_1,1-\delta_2,1$ ($\delta_1$ and $\delta_2$ are introduced below). The gluing does not match $(k+1)$th $U$-derivatives so these will be discontinuous at $U=U_0$. Such discontinuities propagate along characteristic surfaces in spacetime. Hence the solution will exhibit discontinuities in $(k+1)$th derivatives along the surfaces $V=0,\delta_1,1-\delta_2,1$ and $U=U_0$. Where these intersect the origin $r=0$ the solution can become more singular. Consider a spacelike hypersurface that extends to $r=0$ in the Minkowski region of the spacetime, where $r=0$ is smooth. The data on this surface is $C^k$ and piecewise $C^{k+1}$, with bounded discontinuities in $(k+1)$th derivatives. This implies that it is in the Sobolev space $H^{k+1}$. The Cauchy stability argument of \cite{Kehle:2022uvc} implies that the spacetime admits a regular origin to the past of the Minkowski region. Propagation of regularity implies that the solution is in $H^{k+1}$ everywhere including the origin. Sobolev embedding then implies that it is $C^{k-2}$ everywhere including the origin. 

\subsection{Ansatz for the gluing functions}

The functions $P(V)$ and $Q(V)$ in the Ansatz \eqref{eq:PhiAnsatz}, \eqref{eq:BAnsatz} for the gluing functions enforce the boundary conditions for $\Phi$ and $\mathcal{B}$ on $U=U_0$. These are $\Phi=0$, $\partial_V^j\Phi=0$, $\partial_V^j\mathcal{B}=0$ ($1\leq j \leq k$) at $V=0$ and $V=1$, as well as $\mathcal{B}=0$ at $V=0$ and $\mathcal{B}=-\frac{1}{3}\log 2$ at $V=1$. For $V \in [0,1]$ we define $Q(V)=I_V(5,5)$ where $I_V(a,b)$ is the regularized incomplete beta function, defined by $I_x(a,b)=\mathrm{B}(x;a,b)/\mathrm{B}(a,b)$, which we use for the property $\frac{\mathrm{d}}{\mathrm{d}x}\mathrm{B}(x;a,b)=x^{a-1}(1-x)^{b-1}$. In this case, we have $I_V(5,5)=V^5(126-420V+540V^2-315V^3+70V^4)$. The profile function $P(V)$ is defined piecewise as:
\begin{equation}
P(V) = \begin{cases}
    Q(V/\delta_1) & 0 \leq V \leq \delta_1\\
    1 & \delta_1 <V<1-\delta_2 \\
    Q((1-V)/\delta_2) & 1-\delta_2 \leq V\leq 1
\end{cases}
\end{equation}
where the parameters $0<\delta_1<1$, $0<\delta_2\leq1-\delta_1$ are chosen to be $(\delta_1,\delta_2)=(0.1,0.1)$ for our solutions of type (i), and $(\delta_1,\delta_2)=(0.05,0.1)$ for our solutions of type (ii). $P$, $Q$ can be extended to all of $\mathcal{C}$ by defining $P=Q=0$ for $V<0$ and $P=0$, $Q=1$ for $V>1$. The functions $P,Q$ then have the differentiability properties asserted in the main text. 

\subsection{Plots for transverse derivatives of metric variables}

This section shows plots of $\partial_U r$, $\partial^j_U \mathrm{Re}\,\Phi$, $\partial^j_U \mathrm{Im}\,\Phi$, and $\mathcal{B}^j$, where $j=1,2$ for our solution of type (i) and $j=1,2,3,4$ for our solution of type (ii). In these plots we have multiplied by appropriate powers of $r_+$ to render the result dimensionless, noting that we have defined $V$ to be dimensionless so $U$ has dimensions of length squared.

\begin{figure*}[t]
    \centering
    \includegraphics[width=0.9\linewidth]{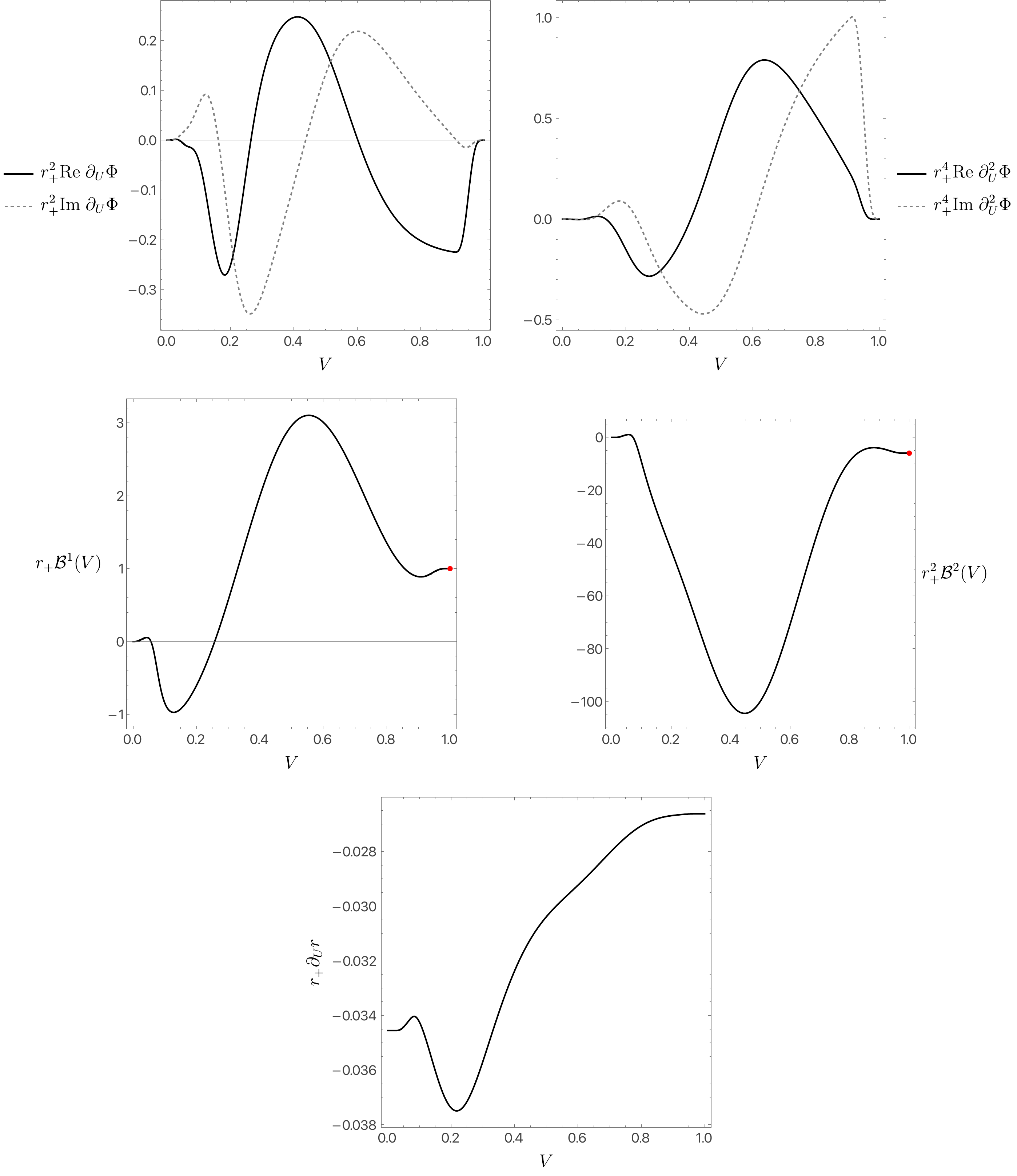}
    \caption{Type (i) solution: plots of $U$-derivatives of $\Phi$, the fields $\mathcal{B}^j$, and $\partial_U r$. The $\Phi$ derivatives stay small, while $\mathcal{B}^j$ shows larger variations; red dots indicate EMP matching values.}
    \label{fig:type_I_appendix}
\end{figure*}

Let us begin with type (i). Fig.~\ref{fig:type_I_appendix} displays the curves for $\partial_U^j \mathrm{Re}\,\Phi$ and $\partial_U^j \mathrm{Im}\,\Phi$ together for each derivative order $j$, alongside $\mathcal{B}^1$, $\mathcal{B}^2$, and $\partial_U r$. While the $U$-derivatives of $\Phi$ remain relatively small, the $\mathcal{B}^j$ exhibit larger variations. In both cases, the solutions show moderate $V$-gradients throughout the integration domain. We also note that, although $\partial_U r$ is negative (as expected), it is not monotonic in $V$. 

\begin{figure*}[t]
    \centering
    \includegraphics[width=0.9\linewidth]{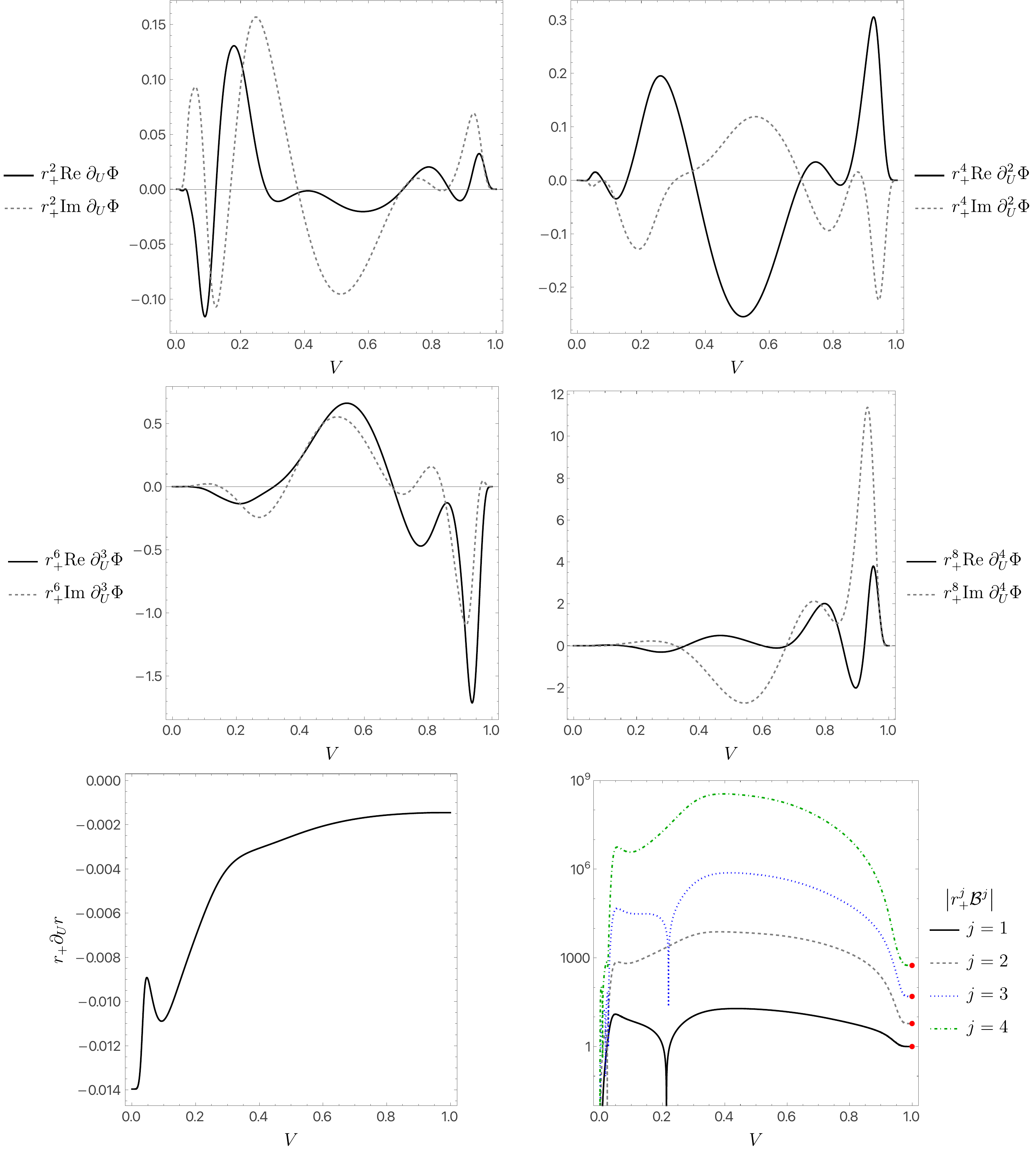}
    \caption{Type (ii) solution: plots of $U$-derivatives of $\Phi$ and the fields $\mathcal{B}^j$. The $|\mathcal{B}^j|$ magnitudes are shown on a logarithmic scale and reach very large values; red disks denote EMP matching values.}
    \label{fig:type_II_appendix}
\end{figure*}

We now turn to the type (ii) solution, shown in Fig.~\ref{fig:type_II_appendix}. For each derivative order $j=1,2,3,4$, the curves for $\partial_U^j \mathrm{Re}\,\Phi$ and $\partial_U^j \mathrm{Im}\,\Phi$ are displayed together. Finally, the magnitudes $|\mathcal{B}^j|$ for all $j=1,2,3,4$ are shown on a single logarithmic plot, which makes all curves visible despite the large differences in scale between the components. In this solution, large $V$-gradients appear throughout the integration domain, which made it considerably harder to obtain numerically. Adding to the challenge, the enormous values of $|\mathcal{B}^j|$, which result from the very small value of $\left.\partial_U r\right|_{V=1}$, posed significant numerical difficulties. It is unclear whether alternative solutions with smaller gradients exist. 

\subsection{The Newton-Raphson search}
For the final step of our numerical algorithm, we employ an extended‑precision quasi‑Newton solver based on the good Broyden method. At this stage, we must select a set of $3k+3$ parameters to tune against the $3k+3$ gluing conditions. This selection was made by inspection and differs between the type (i) and type (ii) constructions.

For the type (i) case, we adjust $\omega_{\chi_0}$, $\varphi_{\chi_0}$, the two components of $\varphi_{\vec{\chi}}$, the overall magnitude of $\varphi_{\vec{\kappa}}$, $\beta_{\chi_0}$, and the three components of $\beta_{\vec{\chi}}$.

For the type (ii) case, we adjust $\omega_{\chi_0}$, the four components of $\omega_{\vec{\chi}}$, $\varphi_{\chi_0}$, the first five components of $\varphi_{\vec{\chi}}$, $\beta_{\chi_0}$, and the first three components of $\beta_{\vec{\chi}}$.

As mentioned in the main text, all data is available in the GitHub repository (\href{https://github.com/jorgealberich/Violation-of-the-third-law-of-black-hole-mechanics-in-vacuum-gravity}{GitHub}).

\subsection{Convergence plots}

At the final stage of our numerical scheme, we perform a quasi-Newton line search over a restricted subset of the parameters controlling the solution. As discussed in the main text, each $C^k$ gluing condition introduces $3k+3$ parameters. To carry out the numerical integrations along the $V$-direction, we employ a spectral-element discontinuous Galerkin (DG) method. The computational domain is subdivided into $N_e$ elements whose boundaries are placed according to a Gauss-Lobatto-Chebyshev grid, thereby clustering the elements near the boundaries $V=0$ and $V=1$. Within each element $\Omega_i$, we use $n_i$ Gauss-Legendre-Lobatto collocation points. For simplicity we take all elements to have the same number of points, so that $n_i=n+1$ (so that $n$ is the order of the polynomial used inside each element). After each quasi-Newton iteration, we are left with $3k+3$ parameters, which we group as a $\mathbb{R}^{3k+3}$ vector $\vec{z}$ as well as $6k+2$ functions which we group into a $\mathbb{R}^{6k+2}$ vector $\mathbf{Q}(V)$, so that
\begin{equation}
\mathbf{Q}(V)\equiv\{r,J,\partial_U^j r,\partial_U^{j-1}\mathcal{A}_U,\partial^j_U \mathcal{B},\partial^j_U {\rm Re}\,\Phi,\partial^j_U {\rm Im}\,\Phi,\partial_U^j \Omega\}\quad\text{with}\quad j=1,\dots,k\,.
\end{equation}

For each choice of resolution $(N_e,n)$, we terminate the Newton iteration once the residuals fall below a prescribed tolerance, which we take to be $10^{-20}$. This yields numerical solutions $\vec{z}^{\,N_e,n}$ and $\mathbf{Q}^{\,N_e,n}$. To assess the convergence of the numerical method, we examine how these quantities change as the resolution parameters $N_e$ and $n$ are varied.

For the vector of functions $\mathbf{Q}^{\,N_e,n}(V) = \big\{Q^{\,N_e,n}_1(V), \dots, Q^{\,N_e,n}_{6k+2}(V)\big\}$, we quantify the difference with a higher-resolution reference solution $\mathbf{Q}^{\,N_e',n'}(V)$ by defining
\begin{equation}
f_{N_e\,n} = \Bigg( \sum_{\alpha=1}^{6k+2} \Big\| Q_\alpha^{\,N_e,n}(V) - \Pi^{(N_e,n)} Q_\alpha^{\,N_e',n'}(V) \Big\|_{L^2}^2 \Bigg)^{1/2}\,,
\end{equation}
where we assume that $N_e'$ and $n'$ are functions of $N_e$ and $n$. A common choice is $N_e'=N_e+z_e$ and/or $n'=n+z$ for fixed $z_e$ and $z$. Here $\Pi^{(N_e,n)}$ denotes the $L^2$-projection of each component of the higher-resolution solution onto the DG space with resolution $(N_e,n)$, defined by
\begin{equation}
\int_\Omega \big(q - \Pi^{(N_e,n)}q\big)\,v_h\,{\rm d}V = 0
\quad\text{for all } v_h \text{ in the DG space}.
\end{equation}
The construction and approximation properties of this projection in the DG context are discussed in standard references on DG methods, e.g., \cite{Riviere:2008}. The integrals are evaluated using the standard DG quadrature on each element, ensuring that the computed difference is measured consistently in the coarse DG space, independent of the fine-grid discretization.

For $h$-refinement, with $N_e' = N_e + z_e$ and fixed $n = n'$, we expect
\begin{equation}
f_{N_e\,n} \sim N_e^{-\min(\tilde{k}+1,n+1)},
\label{eq:dec1a}
\end{equation}
as predicted by standard discontinuous Galerkin convergence theory
\cite{Cockburn:2001,Cockburn:2000,Hesthaven:2008,Riviere:2008}
for a piecewise $C^{\tilde{k}-1}$ continuous \emph{source} function. In our case, the source terms are constructed from piecewise $C^{\tilde{k}}$ functions and their first derivatives, and are therefore piecewise $C^{\tilde{k}-1}$. Although the points of reduced smoothness are not exactly aligned with element interfaces, they lie very close to the domain boundaries. Since the spectral-element mesh clusters elements near $V=0$ and $V=1$, the elements containing these nonsmooth points are parametrically smaller than the bulk elements. As a result, their contribution to the global $L^2$ error is subdominant, and the observed convergence rate effectively corresponds to that of a piecewise $C^{\tilde{k}+1}$ solution, enhancing the decay in Eq.~(\ref{eq:dec1a}) to
\begin{equation}
f_{N_e\,n} \sim N_e^{-\min(\tilde{k}+2,n+1)}\,.
\label{eq:dec1}
\end{equation}

Similarly, for $p$-refinement, with $n' = n + z$ and fixed $N_e = N_e'$, we expect
\begin{equation}
f_{N_e\,n} \sim e^{-\zeta\,n},
\label{eq:dec2}
\end{equation}
with $\zeta$ a constant. One might wonder why the limited regularity of the source terms does not prevent the observed exponential (or spectrally accelerated) convergence with increasing polynomial degree. The elements containing the points at which the source terms lose smoothness are confined to a narrow region near the domain boundaries and shrink rapidly as the mesh is refined due to the Chebyshev clustering. Polynomial approximation on these elements therefore achieves exponentially fast or algebraically enhanced convergence with increasing polynomial degree, and their contribution to the global $L^2$ error remains subdominant. Consequently, the observed convergence under $p$-refinement is governed by the regularity of the solution on the bulk elements, leading to an apparently enhanced decay with increasing polynomial order.

We also define an analogous measure for the parameter vector $\vec{z}^{\,N_e,n}$ by
\begin{equation}
g_{N_e\,n}
=\left\|\mathbf{1}
-\frac{\vec{z}^{\,N_e,n}}{\vec{z}^{\,N_e',n'}}\right\|_{\infty},
\end{equation}
where the division is understood componentwise and $\|\cdot\|_\infty$ denotes the maximum norm. The quantity $g_{N_e\,n}$ therefore measures the relative change in the solution parameters with respect to a higher-resolution reference solution. Its convergence properties mirror those of $f_{N_e\,n}$.

The solution of type (ii), corresponding to $C^4$ gluing, is the most challenging to obtain, and it is therefore the case for which we present convergence tests in the supplemental material. (Naturally, we have also produced these plots for the type (i) solution, but we refrain from presenting them here.)

\begin{figure*}[t]
    \centering
    \includegraphics[width=0.9\linewidth]{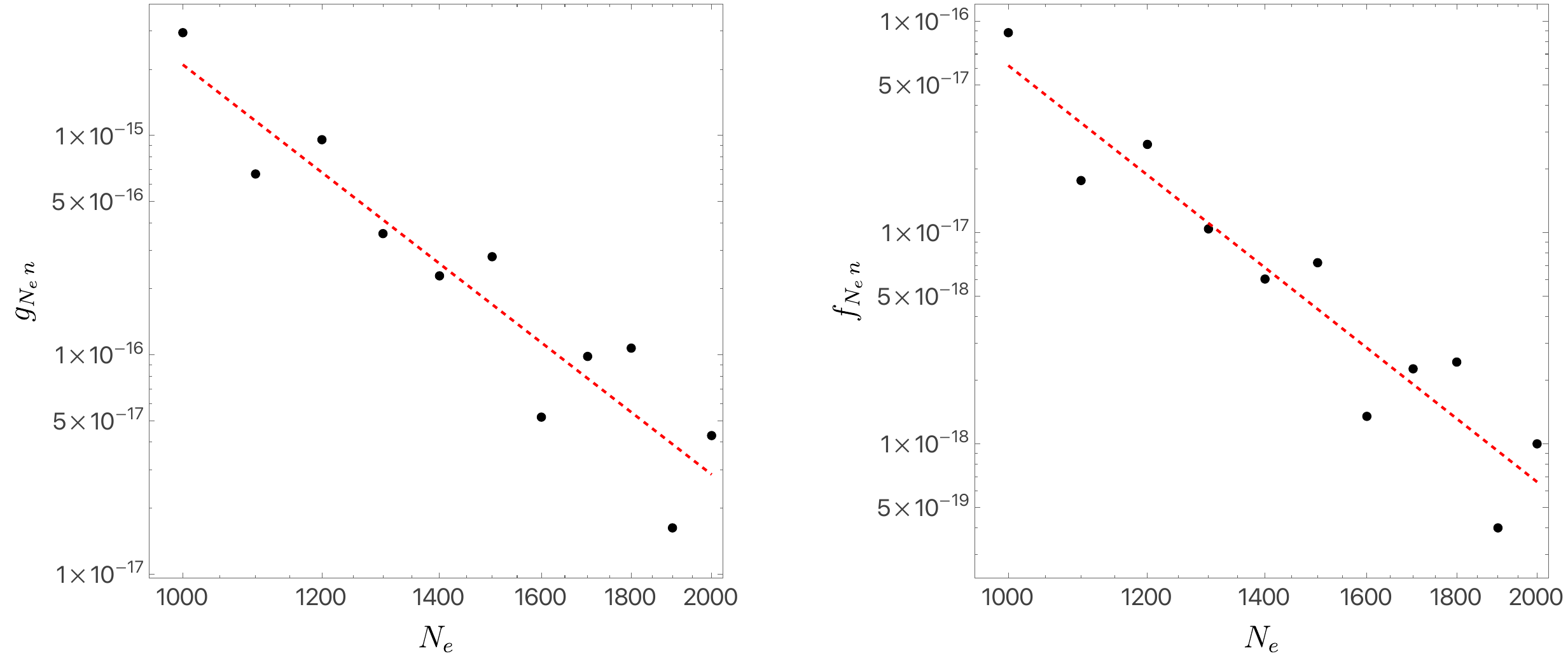}
    \caption{Log–log plots of $g_{N_e,n}$ (left) and $f_{N_e,n}$ (right) as functions of $N_e$. Red dashed lines show power-law fits, yielding exponents consistent with Eq.~(\ref{eq:dec1}).}
    \label{fig:conNe1}
\end{figure*}

In Fig.~\ref{fig:conNe1}, we present $\log$-$\log$ plots of $g_{N_e\,n}$ (left panel) and
$f_{N_e\,n}$ (right panel). The data are shown for $N_e' = N_e + 100$, with
$N_e = 1000, 1100, \ldots, 2000$, while keeping $n = n' = 20$ fixed. The red dashed
line represents a fit to the power-law form $a_0 N_e^{-\gamma}$, where $a_0$ and
$\gamma$ are fitting parameters. From these fits, we extract a scaling exponent
$\gamma \approx 6.20527$ for $g_{N_e\,n}$ and $\gamma \approx 6.54804$ for $f_{N_e\,n}$, both of which are consistent with the prediction of Eq.~(\ref{eq:dec1}).

\begin{figure*}[t]
    \centering
    \includegraphics[width=0.9\linewidth]{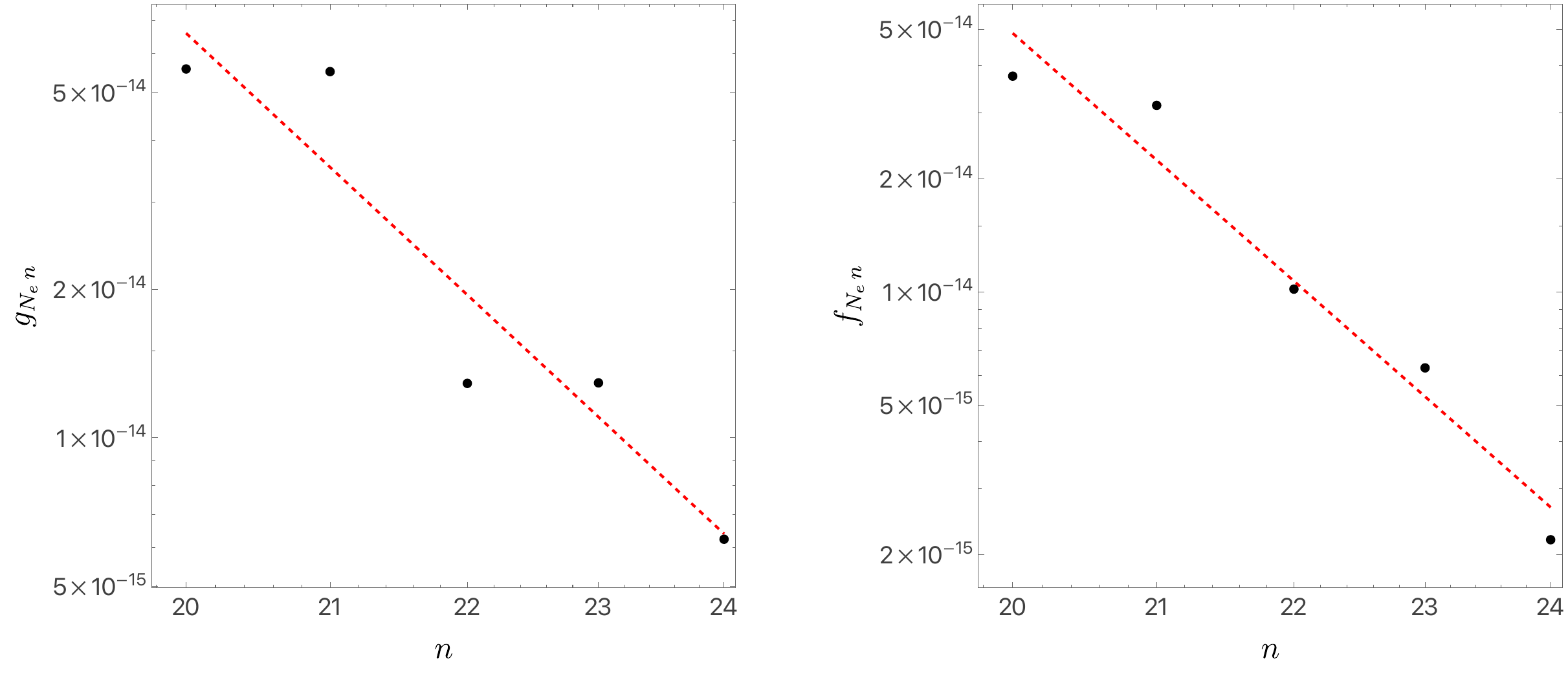}
    \caption{Log plots of $g_{N_e,n}$ (left) and $f_{N_e,n}$ (right) as functions of $n$ at fixed system size. Red dashed lines indicate exponential fits consistent with Eq.~(\ref{eq:dec2}).}
    \label{fig:connp1}
\end{figure*}

Figure~\ref{fig:connp1} shows $\log$ plots of $g_{N_e\,n}$ (left panel) and
$f_{N_e\,n}$ (right panel) as functions of $n$. The results correspond to
$n' = n + 1$, with $n = 20, 21, \ldots, 24$, while the system size is fixed at
$N_e = N_e' = 1000$. The red dashed curve denotes a power-law fit of the form
$a_0 e^{-\gamma\,n}$, where $a_0$ and $\gamma$ are fitting parameters. The extracted
exponents are $\gamma \approx 0.58376$ for $g_{N_e\,n}$ and $\gamma \approx 0.545154$ for $f_{N_e\,n}$, both of which are consistent with the scaling behavior predicted by
Eq.~(\ref{eq:dec2}).

\begin{thebibliography}{99}

\bibitem{Bardeen:1973gs}
J.~M.~Bardeen, B.~Carter and S.~W.~Hawking,
Commun. Math. Phys. \textbf{31}, 161-170 (1973)
doi:10.1007/BF01645742

\bibitem{Israel:1986gqz}
W.~Israel,
Phys. Rev. Lett. \textbf{57}, no.4, 397 (1986)
doi:10.1103/PhysRevLett.57.397

\bibitem{shells}
V. de la Cruz and W. Israel, 
Il Nuovo Cimento A (1965-1970) 51 (1967), p744
https://doi.org/10.1007/BF02721742,
K. Kuchar, 
Czech. J. Phys. B 18 (1968), p435 
https://doi.org/10.1007/BF01698208,
D.~G.~Boulware,
Phys. Rev. D \textbf{8}, no.8, 2363 (1973)
doi:10.1103/PhysRevD.8.2363


\bibitem{Kehle:2022uvc}
C.~Kehle and R.~Unger,
J. Eur. Math. Soc. (2025) doi:10.4171/JEMS/1591
[arXiv:2211.15742 [gr-qc]].

\bibitem{Kehle:2024vyt}
C.~Kehle and R.~Unger,
[arXiv:2402.10190 [gr-qc]].

\bibitem{Reall:2024njy}
H.~S.~Reall,
Phys. Rev. D \textbf{110}, no.12, 124059 (2024)
doi:10.1103/PhysRevD.110.124059
[arXiv:2410.11956 [gr-qc]].
A.~M.~McSharry and H.~S.~Reall,
Phys. Rev. D \textbf{112}, no.10, 104009 (2025)
doi:10.1103/s44z-rbzx
[arXiv:2507.06870 [gr-qc]].

\bibitem{Kehle:2023eni}
C.~Kehle and R.~Unger,
Adv. Math. \textbf{452}, 109816 (2024)
doi:10.1016/j.aim.2024.109816
[arXiv:2304.08455 [gr-qc]].

\bibitem{Myers:1986un}
R.~C.~Myers and M.~J.~Perry,
Annals Phys. \textbf{172}, 304 (1986)
doi:10.1016/0003-4916(86)90186-7

\bibitem{Bizon:2005cp}
P.~Bizon, T.~Chmaj and B.~G.~Schmidt,
Phys. Rev. Lett. \textbf{95}, 071102 (2005)
doi:10.1103/PhysRevLett.95.071102
[arXiv:gr-qc/0506074 [gr-qc]].

\bibitem{Dafermos:2005nh}
M.~Dafermos and G.~Holzegel,
Adv. Theor. Math. Phys. \textbf{10}, no.4, 503-523 (2006)
doi:10.4310/ATMP.2006.v10.n4.a2
[arXiv:gr-qc/0510051 [gr-qc]].

\bibitem{gluing}
S.~Aretakis, S.~Czimek and I.~Rodnianski,
Duke Math. J. \textbf{174}, no.2, 355-402 (2025)
doi:10.1215/00127094-2024-0030
[arXiv:2107.02441 [gr-qc]],
Annales Henri Poincare \textbf{25}, no.6, 3081-3205 (2024)
doi:10.1007/s00023-023-01394-y
[arXiv:2107.02449 [gr-qc]],
Commun. Math. Phys. \textbf{403}, no.1, 275-327 (2023)
doi:10.1007/s00220-023-04800-y
[arXiv:2107.02456 [gr-qc]].

%
\bibitem{Gadioux:2025unn}
M.~Gadioux, H.~S.~Reall and J.~E.~Santos,
[arXiv:2512.10008 [gr-qc]].

\bibitem{characteristic}
A.~D.~Rendall, Proc. R. Soc. Lond. A427, 221 (1990) 
doi:10.1098/rspa.1990.0009,
J.~Luk, Int. Math. Res. Not., Vol. 2012, No. 20, pp. 4625–4678 (2012) doi:10.1093/imrn/rnr201

\bibitem{Kingma:2015}
D.~P.~Kingma and J.~Ba,
Proc. Int. Conf. Learn. Represent. (ICLR) (2015)
doi:10.48550/arXiv.1412.6980

\bibitem{BFGS:1970}
C.~G.~Broyden,
IMA J. Appl. Math. \textbf{6}, 76--90 (1970);
R.~Fletcher,
Comput. J. \textbf{13}, 317--322 (1970);
D.~Goldfarb,
Math. Comput. \textbf{24}, 23--26 (1970);
D.~F.~Shanno,
Math. Comput. \textbf{24}, 647--656 (1970).

\bibitem{Riviere:2008}
B.~Rivi{\`e}re,
Frontiers in Applied Mathematics, Vol.\ 35,
Society for Industrial and Applied Mathematics (SIAM), Philadelphia, PA, 2008.
ISBN: 978-0-89871-656-6.

\bibitem{Cockburn:2000}
B.~Cockburn, G.~E.~Karniadakis, and C.-W.~Shu (eds.),
Lecture Notes in Comput. Sci. Eng., Vol. 11, Springer-Verlag, Berlin (2000)
ISBN: 978-3-540-67221-1

\bibitem{Cockburn:2001}
B.~Cockburn and C.-W.~Shu,
J. Comput. Appl. Math. \textbf{128}, 187--208 (2001)
doi:10.1016/S0377-0427(01)00423-0

\bibitem{Hesthaven:2008}
J.~S.~Hesthaven and T.~Warburton,
Texts in Appl. Math., Vol. 56, Springer, New York (2008)
doi:10.1007/978-0-387-72067-8
ISBN: 978-0-387-72067-8

\end{thebibliography}
\end{document}